\documentclass[
twocolumn,
pra,showpacs,preprintnumbers,amsmath,amssymb]{revtex4}

\usepackage{graphicx}
\usepackage{amssymb}
\usepackage{dsfont}
\usepackage{bm}

\begin{document}

\title{Conditions for one-dimensional supersonic flow of 
quantum gases}

\author{S. Giovanazzi$^{1}$}
\author{C. Farrell$^{1}$}
\author{T. Kiss$^{2}$}
\author{U. Leonhardt$^{1}$}

\affiliation{$^{1}$School of Physics and Astronomy, University of St
  Andrews, North Haugh, St Andrews, KY16 9SS, Scotland }
\affiliation{$^{2}$Research Institute for Solid State Physics and Optics,\\ H-1525 Budapest, P.~O.~Box 49,  Hungary} 
\begin{abstract}
One can use transsonic Bose-Einstein condensates of alkali atoms
to establish the 
laboratory analog of the event horizon and to measure the acoustic
version of Hawking radiation.
We determine the conditions for supersonic flow and the 
Hawking temperature for realistic condensates on
waveguides where an external potential plays the role 
of a supersonic nozzle.
The transition to supersonic speed occurs at the 
potential maximum and the Hawking temperature is
entirely determined by the curvature of the potential.
\end{abstract}


\maketitle

\section{Introduction}

\begin{figure}[t]
\includegraphics[width=8.5cm]{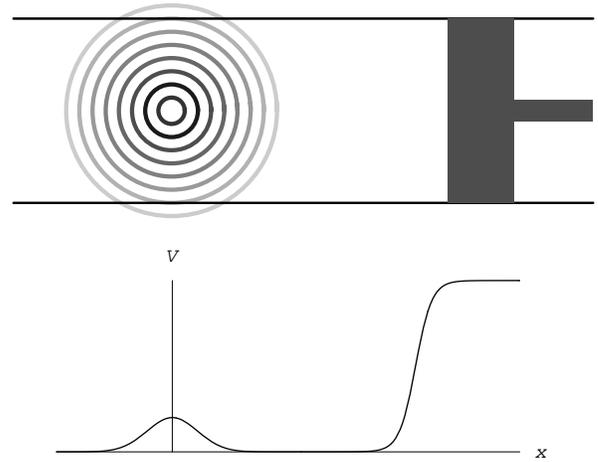}
\caption{\label{fig:piston} 
Scheme of a possible experiment to observe the Hawking effect.
An optical piston pushes a Bose-Einstein condensate,
confined to a waveguide, over a potential barrier.
Both the piston and the barrier can be made by the foci
of blue-detuned light beams acting as the potentials
indicated in the lower part of the figure.
At the barrier the condensate breaks the speed of sound and establishes
the acoustic equivalent of the event horizon.
The sonic analogue of the Hawking effect 
should generate an extra thermal cloud of atoms
where the effective temperature of the cloud depends on the
applied confining potential.
}
\end{figure}

The propagation of sound waves in irrotational fluids is mathematically 
equivalent to wave propagation in General Relativity 
\cite{Unruh,Visser,Bergliaffa}.
This analogy supports an intuitive and simple picture 
for the event horizon \cite{Unruh}:
The horizon is the place where the fluid exceeds the local speed of sound.
One could, in principle, use such a sonic horizon to generate and
measure the acoustic equivalent of the elusive quantum effects
of the event horizon, in particular Hawking radiation 
\cite{Hawking, Brout}.
In practice, the way towards artificial black holes \cite{Book}
has been thorny.
It takes an ultracold quantum fluid to generate a noticeable quantum effect
at the horizon. The only quantum fluids available at the time when
the first idea of artificial black holes appeared in print \cite{Unruh}
were superfluid Helium-4 and Helium-3. 
However, according to the Landau criterion \cite{Tilley},
Helium-4 looses superfluidity well before it reaches the speed of sound,
because Helium-4 is a strongly interacting quantum liquid.
Helium-3  is a more complex quantum liquid
with a wealth of analogies between the physics of its 
elementary excitations and General Relativity or
various other gauge theories \cite{Volovik},
yet so far such analogies have never been experimentally observed
in a direct way.
The advent of alkali Bose-Einstein condensates \cite{PS}
improved the prospects of sonic horizons in simple quantum fluids
and inspired a renewal of interest in their generation 
and their quantum effects 
 \cite{Book,Garay,BS,LKO,LKOReview,FF}.
These condensates are weakly interacting quantum gases,
not primarily quantum liquids, resembling 
very closely the perfect Bogoliubov gas. 
The alkali condensates are the coldest quantum gases 
currently available \cite{Ketterlecold}.
The condensates also allow many ways of experimental manipulation.
For example, condensates can be generated on 
atom chips \cite{Chips} and guided in
current-carrying wires in magnetic fields \cite{Wire}
or light beams \cite{Goerlitz}.
Tightly focused spots of light can be used to manipulate them 
\cite{Shin}, exploiting the dipole force that light exerts on atoms.
Waveguides are advantageous for achieving supersonic flow,
because they can confine condensates to longitudinal areas that are small
enough to prevent the formation of vortices. Otherwise, the turbulence
created would not allow superfluid flow at supersonic speed.
Figure \ref{fig:piston} illustrates schematically a possible set-up
to generate a supersonic flow in a Bose-Einstein condensate.

In this paper we determine the conditions 
required to exceed the speed of sound 
and the resulting Hawking temperature
for condensates on waveguides.
For this, we develop the hydrodynamic theory of the one-dimensional
gas flow through variable longitudinal areas and under the influence
of transversal potentials. 
The one-dimensional gas flow with variable area is a textbook theory
\cite{LL6}.
In this paper we consider a variable potential and arrive at a theory 
that is still simple and fairly general, 
going beyond the immediate concern of Bose-Einstein condensates. 
Having found the velocity and the density profile 
at the sonic horizon, we use the theory of the Hawking effect in fluids
\cite{Visser,LKOReview} 
to compute the Hawking temperature. 
This approach is valid as long as the gas profile varies on longer 
scales than the healing length (the correlation length) 
\cite{LKOReview}.
The same condition justifies the hydrodynamic approximation
\cite{PS} that we use. 
Here the specific properties of the
quantum gas are condensed into an equation of state 
and the dynamics is governed by the equation of continuity
and the Bernoulli equation.  
Calculations of the effective Hawking temperature have been published
before for the one-dimensional gas flow with variable area 
but constant potential \cite{BS}.
However, applying transversal potentials
by tightly focused light beams  \cite{Shin}
seems to be the easiest way to establish a sonic horizon
in a Bose-Einstein condensate.  

For the scheme illustrated in Fig. \ref{fig:piston} 
we find the critical potential
\begin{equation}
\label{eq:r1} 
U_c = m\left(c_0^2 + \frac{v_0^2}{2} - 
\frac{3}{2} \left(v_0 c_0^2\right)^{2/3}\right) \,,
\end{equation}
where $m$ denotes the atomic mass and $c_0$ and $v_0$ are the condensate's initial speed of sound and flow velocity. If the applied potential barrier lies below $U_c$ the quantum gas does not become supersonic. Above $U_c$ the condensate turns from subsonic to supersonic speed at the potential maximum $U_m$. The driving piston will compress the quantum gas such that it always obeys the relation (\ref{eq:r1}) with $U_c=U_m$ where $c_0$ is the local speed of sound immediately in front of the piston and $v_0$ is the flow speed, {\it i.e.} the velocity of the piston.

The more tightly confined the potential barrier is the larger is the resulting velocity gradient at the horizon and the higher is the Hawking temperature $T$ \cite{Unruh,Visser}.  We find
\begin{equation}
\label{eq:r2} 
T= \frac{\hbar\omega_0}{2\pi k_B}\,\frac{\sqrt{3}}{2}
\,,\quad 
m\omega_0^2 =\left. -\frac{\partial^2U}{\partial x^2}\right|_{\rm horizon} \,,
\end{equation}
where $k_B$ denotes Boltzmann's constant. We see that $T$ depends entirely on the curvature of the potential at its maximum and on the atomic mass, constituting the effective frequency $\omega_0$. The numerical factor $\sqrt{3}/2$ is the sole trace of the hydrodynamic properties of the condensate. 

To achieve an optical potential in the order of the condensate's mean-field energy $mc_0^2$ does not pose much of an experimental problem.
The critical issue is the focus required in order to generate a noticeable Hawking effect \cite{Shin}.
For a focus of length $l$, the frequency $\omega_0$ is in the order of $\sqrt{2}\,c_0/l$,
assuming that $m\omega_0^2 \approx 2U_c/l^2 \approx 2mc_0^2/l^2$.
For a narrowly confined sodium condensate with
$c_0\approx 10^{-2}{\rm m}/{\rm s}$ the Hawking energy $k_BT$ reaches about $15{\rm nK}$ if the potential is focused to $l=10^{-6}{\rm m}$.
Such an enhanced thermal cloud could be observable (the record of low temperatures measured so far lies below $1 {\rm nK}$ \cite{Ketterlecold}). 
Since the Hawking temperature is independent of the density,  
see Eq.\ (\ref{eq:r2}), 
one may employ a sufficiently dilute condensate where
inelastic three body losses do not pose a severe limitation
and the condensate's lifetime is long enough.
The piston/barrier scheme sketched in Fig.\ \ref{fig:piston} could act like an evaporative cooling device where the thermal part of the cloud escapes from the subsonic region from the very beginning. 
The Hawking radiation is then the major factor that poses a limit to the final temperature reached.
The dependence of this temperature on the curvature of the potential 
can be exploited to discriminate between a residual thermal cloud 
and the Hawking effect.

\section{Theory}

\subsection{One-dimensional gas flow}

Our theoretical model is based on the concept of the one-dimensional gas flow \cite{LL6}. Here two forces act on the quantum gas. The waveguide confines the condensate to an effective area $A$ and the external potential $U$ acts as a longitudinal force. Consider such a one-dimensional gas flow of particle density $\rho$ and velocity $v$ through the area $A$ with constant discharge $Q$, as expressed in the equation of continuity \cite{LL6}
\begin{equation}
\label{eq:cont} 
\rho v A = Q \,.
\end{equation}
The density $\rho$ and the velocity $v$ are averaged quantities over the area $A$. The stationary gas flow obeys the Bernoulli equation \cite{LL6}
\begin{equation}
\label{eq:bern} 
\frac{v^2}{2} + w = \frac{\mu-U}{m} \,,
\end{equation}
where $\mu$ denotes the chemical potential, the total energy of the gas. 
For the enthalpy $w$ we assume the equation of state
\begin{equation}
\label{eq:w} 
w = G \rho^\alpha \,,\quad G,\alpha > 0 
\end{equation}
that describes a general class of gases, including the ideal gas
and Bose-Einstein condensates within the hydrodynamic approximation
 \cite{PS}.
In the latter case the constants are given by the relations
\begin{equation}
\label{eq:bec} 
\alpha = 1 \,, \quad G = \frac{4\pi\hbar^2 a}{m^2}
\end{equation}
in terms of the (positive) s-wave scattering length $a$ of the condensed atoms
 \cite{PS}.
Both the area $A$ and the potential $U$ may vary along the direction of the gas flow. Equations (\ref{eq:cont}), (\ref{eq:bern}) and (\ref{eq:w}) describe how the gas adjusts to these varying external parameters.

We calculate the local speed of sound, $c$, according to the standard theory of sound waves in fluids \cite{LL6} and find
\begin{equation}
\label{eq:c2} 
c^2 = \rho\, \frac{\partial w}{\partial \rho} = 
G \alpha \rho^\alpha = \alpha w \,.
\end{equation}
It is advantageous to introduce the Mach number
\begin{equation}
\label{eq:nu}
\nu = \frac{v}{c}
= \frac{Q}{\rho c A} =
\frac{Q}{A}\, \frac{\sqrt[\alpha]{G}}{\sqrt{\alpha}}\,
w^{-1/\alpha - 1/2},
\end{equation}
as we obtain from Eqs.\ (\ref{eq:cont}) and (\ref{eq:c2}). 
The relation (\ref{eq:nu}) allows us to express the enthalpy $w$ 
in terms of $\nu$ and, following from Eq.\ (\ref{eq:c2}) 
also the density and local speed of sound, if required,
\begin{eqnarray}
\label{eq:c}
c&=&{\sqrt{\alpha}}
\left(\frac{A}{Q}\, 
\frac{\sqrt{\alpha}}
{\sqrt[\alpha]{G}}\nu\right)^{-\alpha/(2+\alpha)} \,,
\\
\label{eq:rho}
\rho&=&\left(\frac{A}{Q}\, 
\sqrt{\alpha G}\,\nu\right)^{-1/(1+\alpha/2)} 
\,.
\end{eqnarray}
In fact,  within our fluid-mechanical model,
all relevant quantities of the one-dimensional gas flow 
are functions of the Mach number.

\subsection{Supersonic flow}

\begin{figure}[t]
\includegraphics[width=9cm]{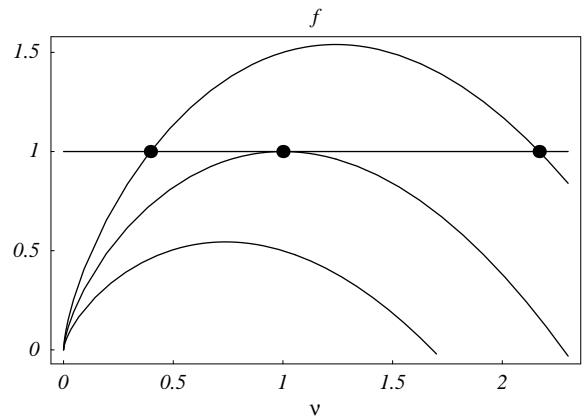}
\caption{\label{fig:f} 
Plot of the Bernoulli function $f(\nu)$ defined in Eq.\ (12)
for Bose-Einstein condensates ($\alpha=1$).
The function is plotted for three $q$ parameters. 
Fluid mechanics implies that $f$ equals unity for a
stationary one-dimensional flow.
The top curve ($q=2$) crosses the line where $f=1$ at two points,
defining a subsonic ($\nu<1$) and a supersonic ($\nu>1$) regime.
The central curve ($q=3/2$) corresponds to the sonic horizon
where the fluid moves with the local speed of sound ($\nu=1$). 
For the lower curve ($q=1$) no stationary flow exists.
}
\end{figure}

Let us establish the conditions for the supersonic flow of a 
one-dimensional gas with the equation of state (\ref{eq:w}).
We divide the Bernoulli equation (\ref{eq:bern})
by $w$ and get
\begin{equation}
\label{eq:ff} 
f(\nu) = 1
\end{equation}
with the function
\begin{equation}
\label{eq:f} 
f = q\, \nu^{\alpha/(1+\alpha/2)} - \frac{\alpha}{2}\, \nu^2 \,,
\end{equation}
see Fig.\ \ref{fig:f}.
All external parameters, in particular the potential $U$ and the area $A$, constitute the single quantity
\begin{equation}
\label{eq:q} 
q = \frac{\mu-U}{m}\, \left(\frac{A}{Q}\, 
\frac{\sqrt{\alpha}}
{\sqrt[\alpha]{G}}\right)^{\alpha/(1+\alpha/2)}
\end{equation}
that may depend on the longitudinal position $x$ along the gas flow. The $q$ parameter is positive, because the total energy $\mu$ is larger than the potential $U$. The exponent $\alpha/(1+\alpha/2)$ of the first term of $f(\nu)$ in Eq.\ (\ref{eq:f}) does not exceed the exponent of $2$ of the second term. Therefore, the function $f(\nu)$ has a maximum that depends on the value of the $q$ parameter. For a critical parameter $q_c$ the maximum of $f(\nu)$ occurs at $f=1$, coinciding with the solution of the scaled Bernoulli equation. To find the maximum, we differentiate $f(\nu)$ with respect to $\nu$ and get 
\begin{equation}
\nu\frac{\partial f}{\partial \nu} =
\frac{\alpha}{1+\alpha/2}\, (f-\nu^2)
\end{equation}
that vanishes at $\nu=\pm 1$ for $f=1$. Consequently, at the critical parameter $q_c$ the gas flows with the local speed of sound, establishing a sonic horizon
\cite{Garay,Unruh,Visser,LKOReview}. In this case the function (\ref{eq:f}) reaches unity at
\begin{equation}
\label{eq:qc} 
q_c = 1 + \frac{\alpha}{2} \,.
\end{equation}
For $q<q_c$ the curve of $f(\nu)$ lies below unity and therefore
no stationary flow exists, whereas for $q>q_c$ the gas 
establishes two solutions, a subsonic and a supersonic regime.
Which one of the two regimes is realized depends on the evolution 
of the flow.
An initially subsonic gas stream stays subsonic until 
the flow reaches the local speed of sound. 
In order to find out how the gas proceeds beyond
the sonic horizon, we expand $q$ and $\nu$ in the 
vicinity of their critical values
\begin{equation}
q = q_c + \delta q \,,\quad \nu = \pm 1 + \delta\nu \,.
\end{equation}
We obtain from the scaled Bernoulli equation (\ref{eq:ff})
with the definitions (\ref{eq:f}) and (\ref{eq:q}),
to lowest order in $\delta q$ and $\delta\nu$,
\begin{equation}
\label{eq:deltaq}
\delta q = \frac{2\alpha}{2+\alpha}\, (\delta\nu)^2 \,.
\end{equation}
Consequently, $q$ reaches a minimum at the critical parameter,
which is consistent with the result that for $q<q_c$
no solution exists.
Near a minimum of the $q$ parameter, $\delta q$
depends quadratically on the distance from the sonic 
horizon, assuming that the second derivative 
of $q$ does not vanish, which is usually the case in practice. 
Therefore, $\delta\nu$ is proportional to the distance.
Consequently, if the flow reaches the local speed of sound
the gas cannot instantly retreat to subsonic speed.
The flow becomes supersonic.
Similarly, a supersonic flow will become subsonic when
$q$ reaches $q_c$.
Sonic horizons are usually transsonic.

\subsection{Horizons}

\begin{figure}[b]
\includegraphics[width=7.5cm]{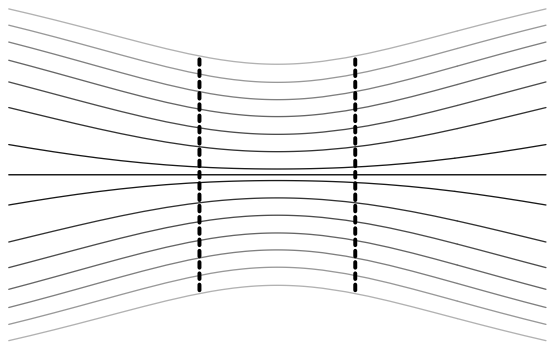}
\caption{\label{fig:gauss} 
A Gaussian light beam may both guide and focus the condensate,
appearing as the optical analogue of the de Laval nozzle.
However, as we have shown, the interplay between longitudinal
confinement and transversal forces will establish two horizons,
if any, {\it i.e.} a natural double de Laval nozzle [6]. 
}
\end{figure}

The flow reaches the speed of sound when the $q$ parameter 
is both minimal and equal to $q_c$.
The latter condition determines how the system parameters
should be adjusted or how they adjust themselves
for stationary transsonic flow.
If the $q$ parameter at the minimum exceeds $q_c$, 
a stationary subsonic flow exists and 
therefore the gas does not become supersonic.
If $q<q_c$ the driving piston compresses the gas such that 
$q$ evolves to reach $q_c$.
The minimum of $q$ depends on the way how the system
parameters vary in Eq.\  (\ref{eq:q}).
If the potential is constant, as in the traditional one-dimensional
gas flow \cite{BS,LL6}, the $q$ parameter is minimal
when the area $A$ reaches a minimum, {\it i.e.} 
at the waist of the nozzle.
Suppose that both the potential $U$ and the area $A$ vary, with
\begin{equation}
\label{eq:u}
U = - \frac{V_0}{A^\beta} \,.
\end{equation}
For example, the intensity of a Gaussian light beam, used to
confine the flowing condensate, is inversely proportional
to the area $A$. Since the optical potential is proportional
to the light intensity we get $\beta = 1$.
We obtain from the requirement that $\partial q /\partial A$
vanishes the critical area 
\begin{equation}
A_c = \left(\frac{\alpha\beta-2\alpha+2\beta}{2\alpha}\,
\frac{V_0}{\mu}\right)^{1/\beta} \,.
\end{equation}
For a Gaussian light beam confining a Bose-Einstein condensate
we get
\begin{equation}
A_c = \frac{V_0}{2\mu} \,.
\end{equation}
Transitions from subsonic to supersonic speed
and vice versa occur at a specific confining area. 
Therefore, a Gaussian beam establishes two
sonic horizons around its waist, if any, see Fig.\ \ref{fig:gauss}.

\subsection{Critical potential}

In the case when the effective confining area $A$ stays 
constant along the gas flow, but the potential varies,
the sonic horizon occurs at the potential maximum $U_m$, 
provided that $q$ can adjust to $q_c$ by changing 
the chemical potential $\mu$ such that
\begin{equation}
\label{eq:connect}
\frac{\mu-U_m}{m}\, \left(\frac{A}{Q}\, 
\frac{\sqrt{\alpha}}
{\sqrt[\alpha]{G}}\right)^{\alpha/(1+\alpha/2)} = 1 +\frac{\alpha}{2} \,,
\end{equation}
as we obtain from Eqs.\ (\ref{eq:q}) and (\ref{eq:qc}).
For example, a driving piston compresses the gas until it 
reaches a stationary flow where it breaks the speed of sound at the 
potential maximum. 
The compression involves changing the energy of the gas, {\it i.e.}
the chemical potential.
In the case the potential barrier is too shallow, 
{\it i.e.} below a critical value $U_c$,
supersonic flow will not occur. 
To give an indication of the required potential height
we calculate how the critical $U_c$ depends on the initial conditions.

Initially, the potential is zero and the gas flows with velocity $v_0$. We read off the chemical potential $\mu$ from the Bernoulli equation (\ref{eq:bern}) and express $\mu$ in terms of the initial speed of sound, $c_0$, and the initial (subsonic) Mach number $\nu_0$,
\begin{equation}
\label{eq:mu}
\mu = mw_0 + \frac{m}{2} v_0^2 = 
\frac{mc_0^2}{\alpha} \left(1+\frac{\alpha}{2}\nu_0^2\right) \,.
\end{equation}
We obtain the initial $q$ parameter by solving the scaled Bernoulli equation (\ref{eq:ff}) for $q$. We express the solution in terms of the chemical potential (\ref{eq:mu}) and get
\begin{equation}
\label{eq:q0}
q_0 = \nu_0^{-\alpha/(1+\alpha/2)}\left(1+\frac{\alpha}{2}\nu_0^2\right)
= \nu_0^{-\alpha/(1+\alpha/2)} \frac{\alpha\mu}{mc_0^2} \,.
\end{equation}
Equation (\ref{eq:q}) implies that $q_c/q_0=(\mu-U_c)/\mu$ for constant $A$, 
which gives
\begin{eqnarray}
U_c &=& \mu\left( 1-\frac{q_c}{q_0}\right) = \frac{mc_0^2}{\alpha}\,\eta \,,
\nonumber\\
\eta &=& 1+\frac{\alpha}{2}\,\nu_0^2 - 
\left(1+\frac{\alpha}{2}\right)
\nu_0^{\alpha/(1+\alpha/2)} \,.
\label{eq:eta}
\end{eqnarray}
For $0\le\nu_0\le1$ we get $1\ge\eta\ge0$ such that the critical potential does not exceed the initial internal energy of the gas $mw_0=\mu-mv_0^2/2=mc_0^2/\alpha$. Formula (\ref{eq:eta}) determines the critical potential (\ref{eq:r1}) in the case of Bose-Einstein condensates where $\alpha = 1$. 

\subsection{Hawking temperature}

The transsonic quantum gas generates the equivalent of Hawking radiation,
a thermal cloud of atoms with the effective temperature \cite{Visser}
\begin{equation}
T= \frac{\hbar\omega_0}{2\pi k_B}
\left. \frac{\partial(v\mp c)}{\partial x}\right|_{\rm horizon} \,,
\end{equation}
where the sign is chosen as the opposite sign of the Mach number at the horizon.
The Hawking temperature thus depends on the gradient of the flow speed
and of the local speed of sound. Both can be expressed in terms of changes in 
the Mach number $\nu$. 
We obtain from Eq.\ (\ref{eq:c}) and the definition of the Mach number
\begin{equation}
\delta c =
-\frac{\alpha c}{2+\alpha}\,\frac{\delta\nu}{\nu} \,,\quad
\delta v = \nu\,\delta c + c\,\delta\nu 
= \frac{2c}{2+\alpha}\,\delta\nu \,.
\end{equation}
Close to the maximum, we represent the potential as
\begin{equation}
U \sim U_m - \frac{m\omega_0^2}{2}\, (\delta x)^2 \,.
\end{equation}
We express $\delta\nu$ in terms of $\delta x$, using the relation (\ref{eq:deltaq})
and the definition (\ref{eq:q}) of the $q$ parameter for constant $A$,
\begin{equation}
(\delta\nu)^2 
= \frac{2+\alpha}{2\alpha}\,
\frac{\omega_0^2}{2}
\left(\frac{A}{Q}\, 
\frac{\sqrt{\alpha}}
{\sqrt[\alpha]{G}}\right)^{\alpha/(1+\alpha/2)}
(\delta x)^2 \,,
\end{equation}
and apply the relationship (\ref{eq:c}) between the local speed of sound 
and the Mach number, which gives
\begin{equation}
\delta v \mp \delta c = \frac{\sqrt{2+\alpha}}{2}\,\omega_0\,\delta x \,.
\end{equation}
In this way we arrive at the Hawking temperature 
\begin{equation}
T= \frac{\hbar\omega_0}{2\pi k_B}\,\frac{\sqrt{2+\alpha}}{2} \,.
\end{equation}
Our result (\ref{eq:r2}) for Bose-Einstein condensates follows for $\alpha=1$. The Hawking temperature is proportional to the characteristic frequency $\omega_0$ that describes the curvature of the potential. ($\omega_0$ is the oscillation frequency of an inverted harmonic oscillator fitted to the potential at the maximum.) The factor $\sqrt{2+\alpha}/2$ depends on the equation of state. No other hydrodynamic properties of the quantum gas contribute to the Hawking temperature.  

\section{Numerical simulation}

\begin{figure}[b]
\includegraphics[width=8.5cm]{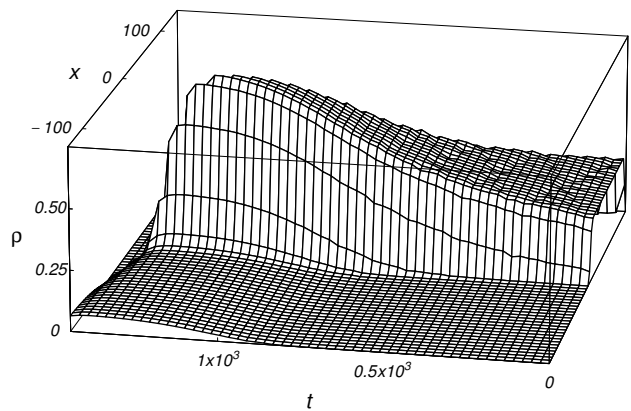}
\caption{\label{fig:simulation} 
Result of the numerical simulation:
Density profile of the evolving condensate
in dimensionless units.
The optical piston compresses the condensate 
and pushes it over the potential barrier,
as indicated in Fig.\ 1.
Here the condensate becomes supersonic
and its density drops dramatically. 
The process continues until the reservoir 
between piston and barrier runs out of atoms.
Our numerical simulations indicate
that the one-dimensional transsonic flow is stable,
{\it i.e.} one-dimensional sonic black holes 
should be observable without being obscured
by instabilities. 
}
\end{figure}

We tested the predictions of our hydrodynamic theory with numerical
simulations of the Gross-Pitaevskii equation \cite{PS} 
\begin{equation}
\label{eq:GP}
i\hbar\frac{\partial\psi}{\partial t} = 
-\frac{\hbar^2}{2m}\,\frac{\partial^2 \psi}{\partial x^2} + 
g_{A}\,|\psi|^2\psi + V\psi
\end{equation}
for the macroscopic wave function $\psi$ of the condensate
averaged over the longitudinal area. 
Here $g_{A}$ refers to the effective s-wave scattering coupling
constant that has been averaged similarly.  
The potential $V$ consists of the sum of two parts, the confining
potential $U$ and the potential of the optical piston
$W$ that is used to drive the condensate from the right to the left
over the potential barrier to supersonic speed.
The condensate is initially confined between the barrier
and the piston.
For the simulations we made the Gross-Pitaevskii equation (\ref{eq:GP}) 
dimensionless such that $\hbar=m=g_{A}=1$, 
by appropriately changing the scales of 
length, time and atomic density.
We used the potentials
$U =\frac{1}{2}\exp\left(-0.125^2x^2\right)$
and
$W=5\left[1+ \frac{1}{2}\tanh(x-x_p-v_p\,t)\right]$
where $x_p$ is the initial position of the piston and $v_p$ is its velocity. 
The initial condensate state at $t=0$ is 
first determined using the
Thomas-Fermi approximation \cite{PS} and 
then propagated in negative imaginary time
in the reservoir between the potential barrier and the piston
in order to find the lowest energy state for the initial potential. 
Finally, it is given a ``kick'' to
match its velocity with the piston speed by multiplying it by a term
$\exp[-iv_p x^2/(2x_p)]$.
We used a perfectly-matched layer 
\cite{Conor}
to simulate the expansion of 
the supersonic gas into empty space on the left edge of the 
computational domain. 
The Gross-Pitaevskii equation is solved via a
Crank-Nicolson discretization and the use of the tridiagonal
matrix algorithm (Thomas algorithm)
\cite{Numerics}.
Figure \ref{fig:simulation} shows the density profile of the evolving
condensate.

\begin{figure}[t]
\includegraphics[width=8.5cm]{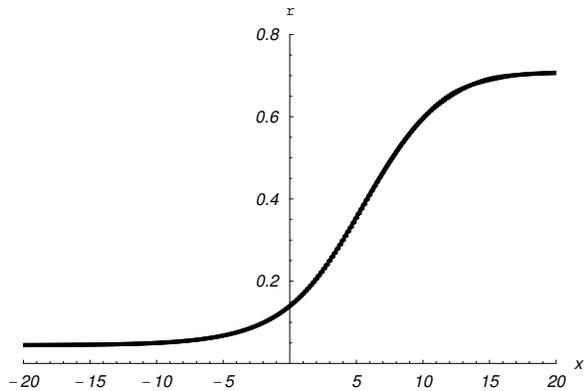}
\caption{\label{fig:num} 
Fit of the theoretical predictions with numerical simulations.
The figure shows the density profile 
in dimensionless units
around the sonic horizon at $x=0$.
The theoretical curve (solid line) is nearly indistinguishable
from the numerical simulation data (points).
In the computation 
we assumed a piston velocity $v_p=0.1$ in dimensionless units
and fitted the chemical potential $\mu$ to the density profile.
A value of $\mu=0.75$ was found to give an excellent fit.}
\end{figure}

When the gas has reached a quasi-stationary regime,
we compared the density profile with our
hydrodynamic theory for stationary flow.
According to this theory, the profile of the Mach number, 
satisfying the relations (\ref{eq:ff}) and (\ref{eq:f}),
depends on the shape of the potential and on two
additional parameters, the chemical potential $\mu$ and the ratio
of the area $A$ and the discharge $Q$. 
Equation (\ref{eq:connect})
connects the parameters and relates them to the maximum 
of the potential barrier. Effectively, only one independent
parameter remains, say the chemical potential $\mu$.
We determined this parameter by fitting the density profile
of the hydrodynamic theory, Eq.\ (\ref{eq:rho}), 
to the numerical simulations
with $\rho=|\psi|^2$ in the quasi-stationary regime.
We found excellent agreement, see Fig.\ \ref{fig:num}.
We also observed that the one-dimensional 
supersonic flow is stable, in agreement
with an earlier theoretical prediction \cite{LKO}.

\newpage

\section{Summary}

We developed a hydrodynamic theory to describe the stationary flow
of a quasi one-dimensional quantum gas.
The gas is subject to an external potential that may vary in longitudinal
direction and is confined to transversal areas that may vary as well, in general.
We determined the general conditions for supersonic flow
and calculated the Hawking temperature of the sonic horizon
for the particular case of a constant area. 
Numerical simulations support our hydrodynamic theory.
Our results indicate that the Hawking effect seems observable
using Bose-Einstein condensates confined to a waveguide.

Our paper was supported by 
the ESF Programme 
Cosmology in the Laboratory,
the Marie Curie Programme of the European Commission,
the Hungarian Scientific Research Fund (contract No.\ T43287), 
the Hungarian Academy of Sciences (Bolyai \"Oszt\"ond\'{\i}j),
the Leverhulme Trust,
and the Engineering and Physical Sciences Research Council.

\end{document}